\documentclass[twocolumn,showpacs,preprintnumbers,amsmath,amssymb
 aps,
 prb,
 lengthcheck,%
]{revtex4-1}
\usepackage{amssymb}
\usepackage{graphicx}
\usepackage{dcolumn}
\usepackage{bm}
\usepackage{hyperref}
\usepackage{makeidx}

\usepackage[english]{babel} 
\usepackage[latin1]{inputenc}
\usepackage{color}
\usepackage{graphicx}
\usepackage{amsmath}

\usepackage{latexsym}
\usepackage{amsthm}

\makeindex
\def\>{\right\rangle}
\def\<{\left\langle}
\def\be{\begin{equation}}
\def\ee{\end{equation}}
\def\ba{\begin{array}{l}}
\def\ea{\end{array}}

\def\be{\begin{equation}}
\def\ee{\end{equation}}
\def\ba{\begin{array}{lll}}
\def\ea{\end{array}}
\def\beq{\begin{eqnarray}}
\def\eeq{\end{eqnarray}}

\newcommand{\cP}{{\cal P}}
\newcommand{\cN}{{\cal N}}
\newcommand{\cC}{{\cal C}}
\newcommand{\cS}{{\cal S}}
\newcommand{\tG}{\tilde{\Gamma}}
\newcommand{\hG}{\hat{\Gamma}}
\newcommand{\hcP}{\hat{{\cal P}}}

\begin{document}

 
\title{Single quasiparticle and electron nano-emitter in the fractional quantum Hall regime}

\author{D. Ferraro$^{1,2}$, J. Rech$^{1}$, T. Jonckheere$^{1}$, and T. Martin$^{1}$ }
\affiliation{$^1$ Aix Marseille Universit\'e, Universit\'e de Toulon, CNRS, CPT, UMR 7332, 13288 Marseille, France\\
$^2$ D\'epartement de Physique Th\'eorique, Universit\'e de Gen\`eve, 24 quai Ernest Ansermet, CH-1211 Geneva, Switzerland}

\date{\today}

\begin{abstract}
We propose a device consisting in an antidot periodically driven in time by a magnetic field as a fractional quantum Hall counterpart of the celebrated mesoscopic capacitor-based single electron source.
We fully characterize the setup as an ideal emitter of individual quasiparticles and electrons into fractional quantum Hall edge channels of the Laughlin sequence. Our treatment relies on a master equation approach and identifies the optimal regime of operation for both types of sources. The quasiparticle/quasihole emission regime involves in practice only two charge states of the antidot, allowing for an analytic treatment. We show the precise quantization of the emitted charge, we determine its optimal working regime, and we compute the phase noise/shot noise crossover as a function of the escape time from the emitter. 
The emission of electrons, which calls for a larger amplitude of the drive, requires a full numerical treatment of the master equations as more quasiparticle charge states are involved. 
Nevertheless, in this case the emission of one electron charge followed by one hole per period can also be achieved, and the overall shape of the noise spectrum is similar to that of the quasiparticle source,  but the presence of additional quasiparticle processes enhances the noise amplitude. 
\end{abstract}

\maketitle
\section{Introduction} 

The on-demand single electron source (SES) based on a driven mesoscopic capacitor, \cite{Feve07} has allowed to achieve interferometric experiments with individual electron and hole wave-packets propagating ballistically along integer quantum Hall (IQH) channels and opened the way to electron quantum optics.\cite{Bocquillon13b} 
It relies on a quantum dot both tunnel-coupled to a quantum Hall edge channel and capacitively coupled to a periodically modulated gate. 
In its optimal regime of operation, this periodic source emits exactly one electron in the first half-period and one hole in the second half-period. \cite{Mahe08, Grenier11} 

This is achieved for an intermediate transparency of the point contact connecting the dot and the edge, when the gate is biased with a square voltage whose amplitude is equal to the dot level spacing.
A complete characterization of this SES can be given in terms of a non-interacting picture in which the electron-electron Coulomb interaction on the dot can be effectively taken into account in terms of a renormalization of the level spacing. This allows to model the action of the mesoscopic capacitor through the Floquet scattering matrix theory. \cite{Moskalets02, Moskalets07, Moskalets08}
 The SES allowed to perform Hanbury Brown and Twiss\cite{HBT} (HBT) experiments with single electrons, as well as Hong-Ou-Mandel\cite{HOM} (HOM) collisions between two electrons propagating on opposite edge channels of the quantum Hall effect. 
 
  \begin{figure}[h]
\centering
\includegraphics[scale=0.50]{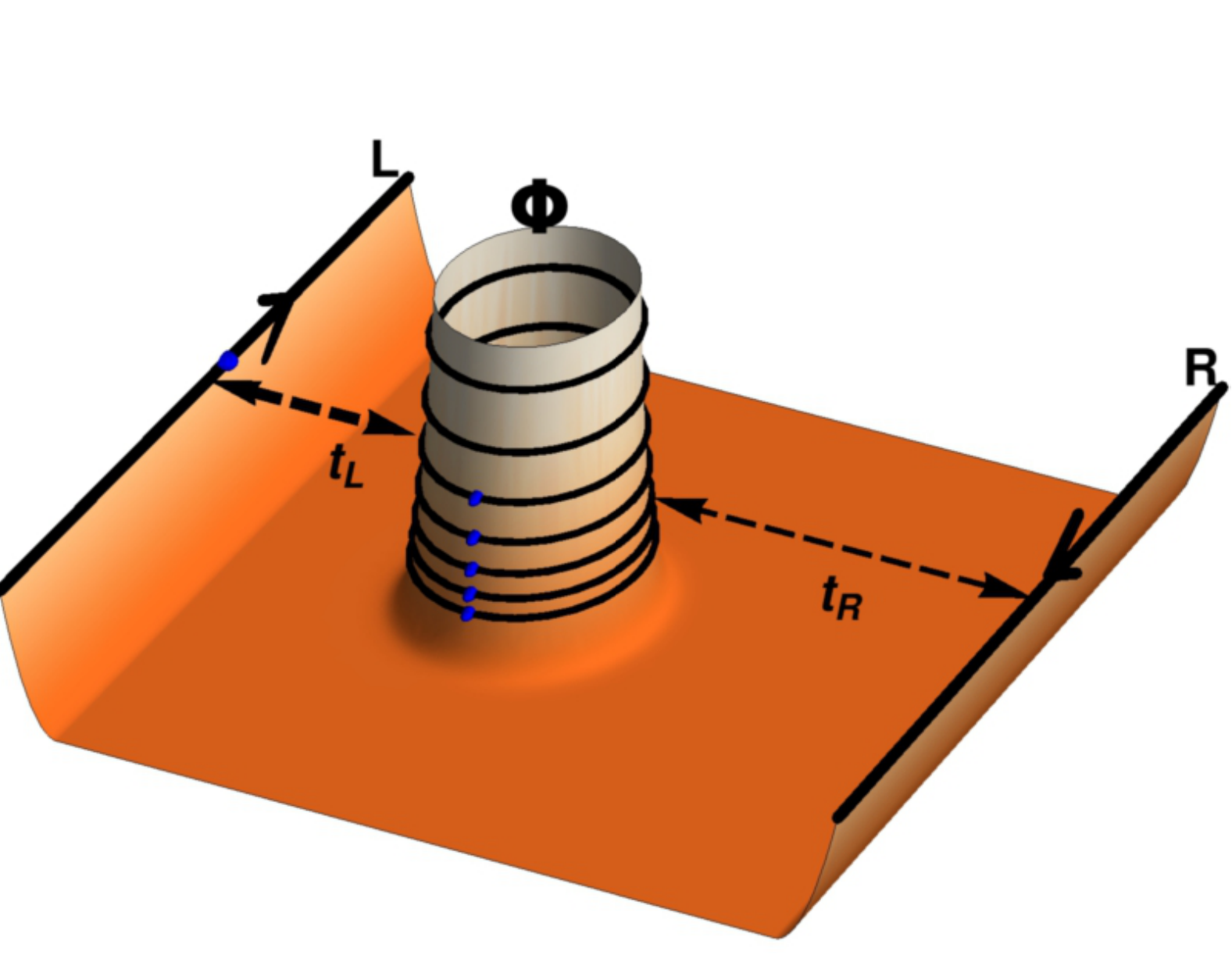}
\caption{Artistic view of a (strongly asymmetric) antidot embedded into an Hall fluid and coupled with edge channels  through tunneling amplitudes $t_{L}$ and $t_{R}$ respectively ($|t_{L}|\gg |t_{R}|$. Dark area represents the Hall fluid while the bright ones are in correspondence of the edges of the Hall bar (black arrows) and of the antidot (with black circles indicating the its energy levels). Blue dots indicated quasiparticles. The antidot is also pierced by a time dependent magnetic flux $\Phi$}
\label{fig1}
\end{figure}

While it is becoming clearer that electron-electron interactions in quantum Hall interferometric devices dramatically affect the nature of the electronic excitations leading to their fractionalization,\cite{Bocquillon13a, Wahl14, Ferraro14} it is natural to contemplate the fascinating possibility of operating such electronic interferometers in the fractional quantum Hall (FQH) regime,\cite{FQHE} where interactions are at their strongest level. There, one would not be dealing with electrons, but rather with emergent Laughlin quasiparticles\cite{Laughlin83} that carry fractional charge, and obey fractional statistics (anyons). \cite{Stern08}
The detection of fractionally charged quasiparticles in the FQH effect has been theoretically predicted in calculations of the DC shot noise characteristics in a tunneling geometry between two counterpropagating edge states.\cite{kane_fisher} Experiments gave a confirmation of these predictions.\cite{glattli_depicciotto}  Alternatively, capacitive measurements using antidot geometries seem to also point toward the detection of fractional charges.\cite{goldman_science}

Unfortunately there is so far no proposal for emitting such single quasiparticles along fractional edge channels: the simple picture of the source as a dot coupled via tunneling to the edge cannot be easily generalized to the case of quasiparticle (QP) and quasihole (QH) emission in the FQH regime. Indeed, for dot transparencies in the optimal regime of emission, the Hall fluid is predicted to be so much depleted that only electrons can tunnel between the dot and the edge of the Hall bar. \cite{Leicht11} Conversely, at higher transparencies, the dot and the edge are strongly coupled so that the output signal reduces to the response of the edge to an applied voltage and is therefore not quantized in general. Granted, in this case, a properly designed Lorentzian voltage pulse could provide an efficient way to realize individual QP injection,\cite{Jonckheere05, Keeling06, Dubois13} at least for the simplest fractional states belonging to the Laughlin sequence. \cite{Laughlin83} 
The main purpose of this work is to characterize the operation of a single Laughlin QP emitter which bears analogies with the driven mesoscopic capacitor source, functioning with a driven antidot which is embedded in the fractional quantum Hall fluid.

An antidot consists of a region of total depletion realized in the Hall fluid (see Fig. \ref{fig1}), and it can be driven either by modulating the Hall magnetic field or with a back gate. As mentioned above, this geometry has been considered as a tool to extract information about both the charge and the statistics of the QP excitations.\cite{goldman_science} In particular, earlier theoretical works investigated the adiabatic pumping of the tunneling amplitudes connecting the antidot with the edges of the Hall bar as a way to emit a perfectly quantized fractional charge per cycle. \cite{Simon00, Das08} In the stationary regime this peculiar geometry also offers the possibility to obtain a persistent current induced by the piercing magnetic field. \cite{Geller97, Geller00} Moreover, noise and higher order current cumulants present features able to disentangle the universal effects associated with the filling factor and the non-universal ones related to the action of the external environment. \cite{Braggio06, Merlo07, Braggio12}  More recently, the same geometry has been discussed in the case of two dimensional topological insulators, \cite{Qi11} where the asymmetry of the antidot configuration turns out to be crucial in order to induce a spin polarized current across the sample. \cite{Dolcetto13} On the experimental side, the periodicity of the conductance peaks in this geometry has been measured as a function of magnetic field and back-gate voltage. This allows to extract charge and exchange statistics of the emergent excitations of the Hall fluid,\cite{Franklin96, Maasilta00, Goldman01, Goldman05} even if these results have been longly debated. \cite{Kane03} 


In this paper, we show that a periodically driven antidot, can either behave as a single-QP source (SQS) or as a SES in the Laughlin sequence of the FQH regime. The setup is shown in Fig.~\ref{fig1} where, in order to ensure the injection of fractionally charged QPs into the left ($L$) edge channel of the Hall bar only, a strong asymmetry is assumed. Here, for working purposes, the periodic drive is obtained by means of a time-dependent modulation of the magnetic field piercing the sample (equivalent modulation of the antidot levels can be achieved with a periodically biased voltage gate). In the present context, the oscillations imposed by the drive constitute only small perturbations with respect to the Hall quantizing magnetic field in order not to deviate from the fractional plateaus. The analysis of this system is carried out using the master equation approach,\cite{Furusaki98} however, in a different manner from what is usually discussed in the literature. Indeed, we need to face two new relevant problems: on the one hand the non-adiabatic time dependence of the drive has to be taken into account, and on the other hand we need to properly characterize the transient regime which is the physically relevant one for our purpose.

By tuning the amplitude of the magnetic field oscillations it is possible to induce the emission into the edge channel of a periodic train either of QPs and QHs, or of electrons and holes. The first case only involves two charge states of the antidot and can be solved analytically. In particular, it is possible to compute the emitted charge per half-period as well as the associated current fluctuations, both these quantities showing remarkable resemblance with the integer quantum Hall (IQH) mesoscopic capacitor setup. \cite{Mahe10, Albert10, Parmentier12} The second case requires a full numerical treatment, due to the large number of charge states involved for the antidot. There, we observe a properly quantized electron charge per half-period, and a vanishing noise at zero frequency.

 However, the noise calculated at the drive frequency exceeds what would be expected for a SES, a consequence of spurious charge emissions randomly occurring during the half-period. All these tunneling processes do not affect the average charge or its fluctuations at zero frequency, but are inherent to the emission process of several quasiparticles in this system. Despite the great number of processes involved, in the regime of electron emission it is also possible to define an effective escape time for the excitations which can be experimentally extracted from the measurement of the first harmonic of the current. 

The paper is divided as follows. In Sec. \ref{Model} we discuss the model of a finite length chiral Luttinger liquid describing the antidot in the Laughlin regime of the FQH effect, coupled to the edges of the same fractional Hall bar. Sec. \ref{Transport} is devoted to the derivation of the tunneling rates (for the master equation approach) as well as the relevant physical quantities, such as the occupation number of the antidot, the current and noise along the edge, which are all essential in order to characterize the performance of the device in both the SQS and the SES regimes. A discussion of the parameters for the optimal emission regime in the two cases is then carried out. In Sec. \ref{SQS} we consider the optimal regime of emission of QPs, where a simple analytical treatment is possible, characterizing the device as a perfect SQS. We investigate the emitted charge, as well as the current fluctuations, showing that the considered setup behaves in exactly the same way as the SES realized in the integer regime, up to the renormalization of the charge of the carriers to account for their fractional nature, $e^{*}= \nu e$. The optimal regime of emission of electrons and holes (SES) is investigated in Sec. \ref{SES}  using a full numerical approach. A perfect quantization of the average emitted charge 

is obtained also in this case, even though additional tunneling processes are required in order to properly describe the system. These can be detected in the finite frequency noise and lead to additional complications in defining the escape time of the electron (hole) from the antidot. In Sec. \ref{sec:estimates}, we perform estimates of the various physical parameters appearing in our SQS calculations in order to compare them with their counterpart in actual experimental realizations in the IQH regime, and we justify the feasibility of our proposed setup. Sec. \ref{Conclusions} is devoted to conclusions, while an Appendix discusses the connection between zero noise contribution at zero frequency and the absence of charge fluctuations during a period. 


\section{Model} \label{Model}

The starting point of our discussion is the antidot geometry proposed in Refs.~\onlinecite{Braggio06, Merlo07}, and schematically illustrated in Fig.~\ref{fig1}. The FQH fluid is depleted by means of an electrostatic gate creating a circular ``empty'' region into the Hall bar. According to Wen's hydrodynamical approach \cite{Wen95} for the description of FQH states belonging to the Laughlin sequence \cite{Laughlin83} with filling factor $\nu=1/(2n+1)$ ($n\in \mathbb{N}$), the Hamiltonian associated with the boundaries (right edge $R$, left edge $L$ and antidot $ad$) is quadratic in terms of the edge-magnetoplasmon creation ($a^{\dagger}_{l, s}$) and annihilation ($a_{l,s}$) operators and can be written in the form ($\hbar=1$)
\be
H^{0}_{l}= \epsilon \sum^{+\infty}_{s=1} s a^{\dagger}_{l, s}a_{l,s}+E_{c}^{l} (N_{l}-N_{\Phi})^{2}
\label{H_zero}
\ee 
where $l=R, L, ad$ labels the various elements of the device, $\epsilon= 2 \pi v/L_{l}$ is the energy associated with the plasmonic modes and $E^{l}_{c}= \pi v \nu /L_{l}$ is the charging energy associated with the zero modes, $L_{l}$ being the length of the $l$-th edge. The propagation velocity $v$ along the edges is assumed constant throughout all the setup. 
$N_{l}$ is the number of QPs enclosed by the edge $l$ (with respect to a fixed background $N^{0}_{l}$) and $N_{\Phi}=\Phi/\Phi_{0}$ is the number of elementary flux quanta $\Phi$ of magnetic field piercing the antidot section ($\Phi_{0}=2\pi/|e|$ the elementary flux quantum). Notice that the last term in Eq.~(\ref{H_zero}) is reminiscent of the minimal Aharonov-Bohm coupling $\textbf{j} \cdot \textbf{A}$, with $\textbf{j}$ the current density along the edge of the antidot and $|\textbf{A}|= \Phi/ L$ the vector potential felt by the antidot itself. \cite{Geller97} In the following, we consider a finite length $L_{ad}=L$ only for the antidot with a consequent non-zero energy $E^{ad}_{c}=E_{c}$, while we assume the thermodynamic limit $L_{R}, L_{L}\rightarrow+\infty$ for the $R$ and $L$ edges (thus yielding $E^{R}_{c}, E^{L}_{c}\rightarrow 0$). Under these conditions the zero mode contribution plays a relevant role only for  the dynamics of the antidot, while the other edges are only described in terms of their plasmonic modes.

As long as the Hall fluid is present between the antidot and the boundaries of the Hall bar (see Fig.~\ref{fig1}), the dominant tunneling process involves single-QPs with charge $e^{*} = \nu e$ ($e$ the electron charge). \cite{Kane92} Therefore the local tunneling Hamiltonian connecting the antidot with the edges of the Hall bar is given by 
\be
H^{T}_{j}= v \left[t_{j}\Psi^{\dagger}_{ad}(x_{j}) \Psi_{j}(0)+H. c.\right] \quad \text{with } j= L, R.
\ee 
Here, the tunneling amplitudes $t_{j}$ are related to the overlap between the Laughlin wave-functions \cite{Levkivskyi10} on the different edges and decay exponentially with their distance. Tunneling processes occur at points $x_{j}$ (see Fig.~\ref{fig1}), whose precise location is not relevant as long as the tunneling is assumed to be local. \cite{Chevallier10, Dolcetto12} The vertex operator associated with the annihilation of one single-QP can be written, according to the standard bosonized description,\cite{Miranda03} as 
\be
\Psi_{l}(x)= \frac{1}{\sqrt{ 2\pi \alpha}} e^{i \varphi_{l}(x)} e^{i \pi \nu \frac{x}{L_{l}}}.
\label{Psi}
\ee
The bosonic field $\varphi_{l}(x)$ appearing in the exponent can be naturally decomposed into the sum of a plasmonic ($\varphi^{p}_{l}$) and a zero mode ($\varphi^{0}_{l}$) contribution given respectively by
\beq
\varphi^{p}_{l}(x)&=& \sqrt{\frac{2 \pi \nu }{L_{l}}}\sum_{k_{l}>0}\left(\frac{1}{\sqrt{k_{l}}} a_{l, k_{l}}e^{i k_{l} x} +H.c.\right)e^{-k_{l}\alpha/2}  \nonumber\\
\varphi^{0}_{l}(x)&=&\frac{2 \pi }{L_{l}} \nu N_{l}x-\chi_{l}
\label{modes}
\eeq
with $\alpha$ a finite length cut-off. The operator $\chi_{l}$ satisfies 
\be
[\chi_{l}, N_{l'}]= i \delta_{l, l'}
\ee
and, once exponentiated, plays the role of a Klein factor, which is essential to provide the correct exchange statistical properties between excitations from different edges. \cite{Braggio06, Merlo07, Guyon02} Notice that the last phase factor in Eq.~(\ref{Psi}) has been introduced in order to satisfy the boundary conditions 
\be
\Psi_{l}(x+L_{l})= \Psi_{l}(x) e^{i 2\pi \nu N_{l}}
\ee
counting the number of fractional excitations enclosed by the edge. \cite{Geller97, Geller00}

\section{Transport properties} \label{Transport}

\subsection{Tunneling rates}

To lowest order in the tunneling Hamiltonian the transport properties depend on the tunneling rates which, by exploiting the periodicity associated with the antidot and assuming the plasmon modes fully relaxed to thermal equilibrium, can be written as \cite{Braggio06, Merlo07, Braggio01, Braggio03, Cavaliere04} 
\be
\tilde{ \Gamma}_{j}(E)=\sum_{p=-\infty}^{+\infty} w_{p}\gamma_{j}(E-p\epsilon) ,
\ee
with $\epsilon$ defined in Sec. \ref{Model}.
This combines the standard expression for the tunneling rate at finite temperature and infinite length of the edges \cite{Roddaro03, Martin05}
\begin{align}
\gamma_{j}(\xi) =& |t_{j}|^{2} \frac{\omega_{c}}{(2\pi)^{2}} \left(\frac{\beta \omega_{c}}{2 \pi} \right)^{1-\nu} \nonumber \\
& \qquad \times B\left[\frac{\nu}{2} +i\frac{\beta \xi}{2\pi},  \frac{\nu}{2} -i\frac{\beta \xi}{2\pi} \right]e^{\beta \xi/2},
\label{rate_gamma}
\end{align}
with the correction accounting for the finite length of the antidot\cite{Braggio00}   
\be
w_{p}= \left(\frac{\epsilon}{\omega_{c}} \right)^{\nu}e^{-p \epsilon/\omega_{c}} \frac{\Gamma(\nu +p)}{\Gamma(\nu)p!}\Theta(p+0^{+})
\label{wp}
\ee
evaluated analytically in the very low temperature limit $\beta \epsilon \gg 1$ ($\beta$ being the inverse temperature) and for $\epsilon/\omega_{c} \ll 1$. Here, we introduced the energy cutoff $\omega_{c}=v/\alpha$, and used $B[a, b]$ and $\Gamma(a)$ as Euler's Beta and Gamma functions respectively.
These tunneling rates present a peaked structure near $E/E_{c}\approx 0$, with a maximum value decaying exponentially for negative energy (with a scale set by temperature), and as a power-law for positive one. Similar peaks also appear at $E/E_{c}\approx 2p/\nu$ ($p\in \mathbb{N}^{0}$) as a consequence of plasmonic excitations, only with a less pronounced amplitude due to the damping factor $w_p$.
 
When the tunneling rates are small compared to both the temperature and the charging energy ($\tilde{\Gamma}_{j}<\beta^{-1}<E_{c}$), it is possible to restrict the analysis to the sequential regime, \cite{Beenakker91, Furusaki98} where only single-QP tunneling processes involving one excitation (incoming or outgoing with respect to the antidot) contribute to the dynamics of the system. In this approximation the relevant rates are the transition probabilities between the initial antidot state with $N$ excitations at  time $t=0$ and the final one with $N\pm 1$ excitations at time $t$. 

\subsection{Master equation}

A simple and useful way to generalize this idea involves the master equation approach, which allows to characterize the time evolution of the probability of occupation of the antidot. \cite{Braggio06, Merlo07, Furusaki98, Geller00} In particular, the probability $\mathcal{P}_{N}(t)$ of having a fixed number $N$ of QPs in a strongly asymmetric antidot ($|t_{L}|\gg |t_{R}|$) at a given time $t$ satisfies the first order differential equation
 \begin{align}
\frac{d \cP_N}{d t} =& \sum_{N'} \left[ \tG \left( E^{N' \to N}\right) \cP_{N'} - \tG \left( E^{N \to N'}\right) \cP_{N} \right] , 
\label{master_general}
\end{align}
(dropping the subscript $j$ in $\tilde{\Gamma}_{j}$ because of the strong asymmetry), where the transition energies are given by
\begin{equation}
E^{N \to N'} = E_c \left[ \left( N - N_\phi \right)^2 - \left( N' - N_\phi \right)^2 \right].
\end{equation}
Since we want to tune the number of QPs on the antidot, we allow modifications of $N_\phi$, with respect to a reference value,\cite{Note2} chosen to be a half-integer, so that
\begin{equation}
N_\phi = N_0 + \frac{1}{2} + \delta(t) .
\label{eq:Nphi}
\end{equation}
Setting $n = N-N_0$, $p = N'-N_0$ and introducing  $\hcP_n = \left. \cP_N \right|_{N = n+N_0}$, the master equation can be conveniently rewritten under a matrix form as
\begin{equation}
\frac{d \hcP (t)}{d t} = \hG  \hcP (t)  ,
\label{eq:mastermatrix}
\end{equation}
where $\hcP (t)$ is a column vector whose elements are the occupation probabilities $\hcP_n (t)$, and $\hG$ is a square matrix whose elements are given by
\begin{align}
\hG_{n p} =& \tG \left( -(n-p) (n+p-1-2 \delta(t)) E_c \right) \nonumber\\
&- \left[ \sum_{q \in \mathbb{Z}} \tG \left( (n-q) (n+q-1-2 \delta(t)) E_c \right) \right] \delta_{n p}.
\label{eq:gammamatrix}
\end{align}
In practice, while various possibilities are foreseeable, we will mostly focus on a square drive, defined over one drive period $T$ as $\delta (t) = \delta\times \text{Sgn} \left( \frac{T}{2} - t\right)$.

\subsection{Occupation number, current and noise}

The occupation $\cN (t)$ of the antidot is readily obtained upon summing up the occupation probabilities solution of the master equation weighted by their corresponding number of QPs
\begin{equation}
\cN (t) = \sum_{n \in \mathbb{Z}} n \hcP_n (t).
\label{eq:dot_occ}
\end{equation}
Of course, this occupation is defined with respect to the background reference set by $N_0$ (cf. Eq.~\eqref{eq:Nphi}).

The total charge on the antidot is readily obtained from the occupation as $Q (t) = e^* \cN(t)$. It follows that the current flowing from the antidot to the edge is simply given by
\be
I (t) = - \frac{d Q (t)}{dt} = - e^* \sum_{n \in \mathbb{Z}} \sum_{p \in \mathbb{Z}} n \hG_{n p} \hcP_p (t) .
\ee

While the current provides crucial information on the operation of the antidot source, relevant information can also be obtained from its noise characteristics. Indeed, the finite frequency signal of the current current correlations allows a finer characterization of the operating conditions of the QP or electron emitter, especially with regard to its dependence on the escape time. \cite{Mahe10,Parmentier12}

When the escape time is much smaller than the period of the drive, the antidot emits
QP or QH (alternatively, electrons and holes) in essentially a periodic manner and the main contribution to the noise is due to the uncertainty of the emission time within each period. This constitutes the regime of so-called ``phase noise'', which is due to the random jitter of triggering of the drive and the actual emission time. On the opposite, when the escape time of QP (QH) (alternatively, electrons or holes) is much larger than the period of the drive, nothing guarantees that the escape from the antidot really occurs, and these rare events give rise to a shot noise like contribution. 

The master equation which is employed here allows to describe both regimes as well as the crossover between the two. However, prior noise experiments performed with the mesoscopic capacitor as the emitter tend (for convenience) to measure the noise at a frequency close to that of the drive, and chose instead to modify the escape time by tuning the transmission of the capacitor coupled to the edge, in order to explore the range of parameters.    

We define the current-current correlation as:
\begin{equation}
\cC_I (t, t') = \langle \delta I (t) \delta I (t+t') \rangle ,
\end{equation}
where $\delta I (t) = I(t) - \langle I(t) \rangle$. 

Because of the periodic drive, the current-current correlation $\cC_I (t, t')$ depends on both $t$ and $t'$, and is $T-$periodic in time $t$. Since we are only interested in the behavior with respect to the time difference $t'$, we consider a time-averaged quantity defined as
\begin{equation}
\cS_I (t') = 2  \int_0^T \frac{dt}{T} \langle \delta I (t) \delta I (t+t') \rangle = 2 \overline{\langle \delta I (t) \delta I (t+t') \rangle} ,
\end{equation}
and the corresponding quantity in frequency space
\begin{equation}
\cS_I (\omega) = \int dt' \cS_I (t')  e^{i \omega t'} .
\end{equation}

As it turns out, the simplest way to derive the current noise is to first access the charge noise as the two are trivially related: $\cS_I (\omega) = \omega^2 \cS_Q (\omega)$.\cite{Albert10} Let us first focus on the charge correlation function $\langle Q(t) Q(t+t') \rangle$ and consider for simplicity that $t' > 0$. 
The charge correlation is only finite when the dot is occupied both at time $t$ and at time $t+t'$. The actual value of the charge correlation is obtained by summing over all possible occupations $n_1$ and $n_2$, multiplied by the joint probability of having $n_1$ QPs at time $t$ and $n_2$ QPs at time $t+t'$. \cite{Korotkov94} The latter is further written, using conditional probabilities, as the product of the probability $\hcP (n_1,t)$ of having $n_1$ QPs occupying the antidot at time $t$, and the conditional probability $\hcP(n_2,t+t' | n_1, t)$ of having $n_2$ QPs at time $t+t'$ given that there were $n_1$ at time $t$. One is left with
\begin{equation}
\langle Q(t) Q(t+t') \rangle = {e^*}^2 \sum_{n_1, n_2} n_1 n_2 \hcP(n_1,t) \hcP (n_2,t+t' | n_1, t).
\end{equation}
Numerically, this conditional probability is obtained by propagating the condition $\hcP (n_1,t) = 1$ through time, using the master equation in Eq.~(\ref{eq:mastermatrix}).

Performing the same calculation for negative values of $t'$, and accounting for the average charge, we have
\begin{align}
\cC_Q (t, t') &= \langle \delta Q (t) \delta Q (t+t') \rangle  \nonumber \\
 &= {e^*}^2 \sum_{n_1, n_2} n_1 n_2  
\left\{ \theta(t') \hcP(n_1,t) \Delta \hcP (n_2,t+t' | n_1,t) \right. \nonumber \\
 & + \left. \theta(-t') \hcP(n_2,t+t') \Delta \hcP (n_1,t | n_2,t+t') \right\}
\label{eq:cqtt}
\end{align}
with $\Delta \hcP (n_1,t_1 | n_2,t_2) = \hcP(n_1,t_1 | n_2, t_2) - \hcP(n_1,t_1)$. 
Computing the average over $t$ and taking the Fourier transform, this ultimately leads to the following expression for the frequency-dependent current noise
\begin{align}
\cS_I (\omega) &= 4 {e^*}^2  \omega^2 \int_0^{\infty} dt' \cos \left( \omega t' \right) \int_0^T \frac{dt}{T}  \nonumber \\
&  \times \sum_{n_1, n_2} n_1 n_2 \hcP(n_1,t) \Delta \hcP (n_2,t+t' | n_1,t) .
\end{align}

\subsection{Tuning the drive}

\begin{figure}[htb]
\centering
\includegraphics[scale=0.62]{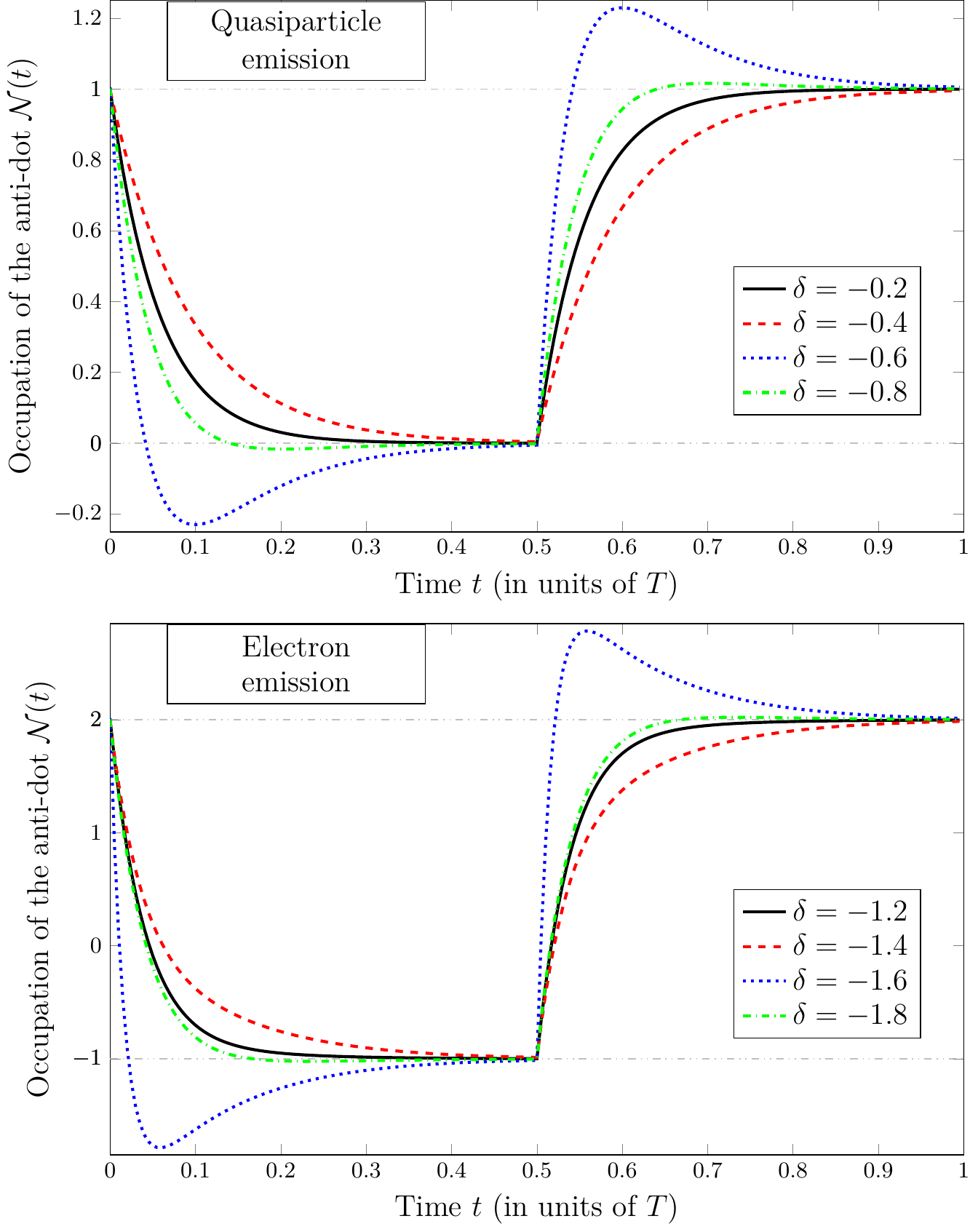}
\caption{Occupation of the antidot over time as a function of the drive amplitude $\delta$ for the SQS (top) and the SES (bottom) regime. Other parameters are: $\nu = 1/3$,  $E_c/\omega_c = 0.01$, $\beta E_c = 20$ and $\gamma_0 T = 100$. }
\label{fig2}
\end{figure}

We now want to determine the optimal regime of operation of the source. 
First, we need to ensure that the antidot has sufficient time to emit/absorb QPs every half-period of the drive. To meet that goal, let us consider the situation of a rather large tunneling rate, say $\gamma_0 T \simeq 100$ where 
\begin{equation}
\gamma_0 =\left( \frac{ 2 E_c}{\nu \omega_c} \right)^\nu  |t_L|^2 \frac{\omega_c}{(2 \pi)^2} \left( \frac{\beta \omega_c}{2 \pi} \right)^{1-\nu} \frac{\Gamma \left( \frac{\nu}{2}\right)^2}{\Gamma(\nu)},
\end{equation}
is the maximum tunneling rate between any two energy levels, and follow the evolution of the dot occupation over time for different values of the drive amplitude $\delta$.
The results for the occupation $\cN (t)$ are presented in Fig.~\ref{fig2} at filling factor $\nu = 1/3$, for a given set of parameters satisfying the constraints of our model, namely $\beta E_c \gg 1$ and $E_c/\omega_c \ll 1$.
These two plots clearly represent the two regimes of interest for the source. The upper panel shows variations of the antidot occupation between 0 and 1, corresponding to the emission/absorption of a single QP. The lower panel shows variations between $-1$ and 2, corresponding to the emission/absorption of an electron charge (in the form of 3 QPs, consistently with the standard picture for the fractional states in the Laughlin sequence). Notice that the negative occupation is an artifact of the choice of the background reference, as the two extrema of the occupation are symmetric with respect to $N_\phi - N_0 = 1/2$.
 The optimal regime of operation corresponds to a situation where the occupation is close to an integer value at the end of every half-period, while behaving monotonously over each half-period. From the results of Fig.~\ref{fig2}, it seems the most appropriate choice for the drive amplitude is $\delta = -0.2$ and $\delta = -1.2$ for the SQS and the SES regime respectively. 

Strikingly, the evolution of the two types of occupation in the upper and lower panel (QP and electrons) bear strong similarities. When the drive amplitude is larger than the optimal drive in both cases, one sees that upon imposing the AC drive the occupation of the antidot overshoots the zero or single QP or electron occupation: for QP (electrons), the trigger of the drive brings the antidot in a configuration which is slightly less than zero, while at the half period, the occupation is on average larger than one.  This overshoot is slightly higher in the electron case than in the QP case. In the optimal case ($\delta = -0.2$ and $\delta = -1.2$ respectively), the response to the drive bears strong similarities with that of the mesoscopic capacitor.  

The remainder of this paper is devoted to the detailed study of these two specific cases. There, we will focus on the "stationary" state, still submitted to a periodic drive but independent of the initial conditions, namely the state of the system when the drive was turned on.

\section{antidot as a SQS} \label{SQS}

We solve numerically the master equation in Eq.~(\ref{master_general}) over several drive periods, considering a vector column $\hcP$ of 31 elements ($n=-15,...,15$) using a matrix version of the fourth-order Runge-Kutta scheme. As it turns out, most occupation probabilities are vanishingly small in this case, and one can focus on a very much reduced set of equations in order to properly describe the behavior of the system, allowing for an analytic treatment.

In practice, one thus only needs to keep track of the probabilities for having a singly occupied or an empty antidot, namely $\hcP_1 (t)$ and $\hcP_0 (t)$ respectively. Focusing on a single half-period, corresponding e.g. to the emission of a QP, their dynamics is captured by the following set of equations
\begin{align}
\frac{d \hcP_1 (t)}{d t} &= -\gamma  \hcP_1 (t)  ,  \\
\frac{d \hcP_0 (t)}{d t} &= \gamma  \hcP_1 (t)  , 
\end{align}
where $\gamma = -\hat{\Gamma}_{11} = \tG (-2  E_c \delta)$. The other half-period, corresponding to the absorption of a QP, is described by similar equations, only exchanging $\hcP_0$ and $\hcP_1$. Note that by definition, $\gamma (\delta =0) = \gamma_0$ and that $\gamma$ then rapidly decreases as one raises $|\delta|$.

\begin{figure}[t]
\centering
\includegraphics[scale=0.62]{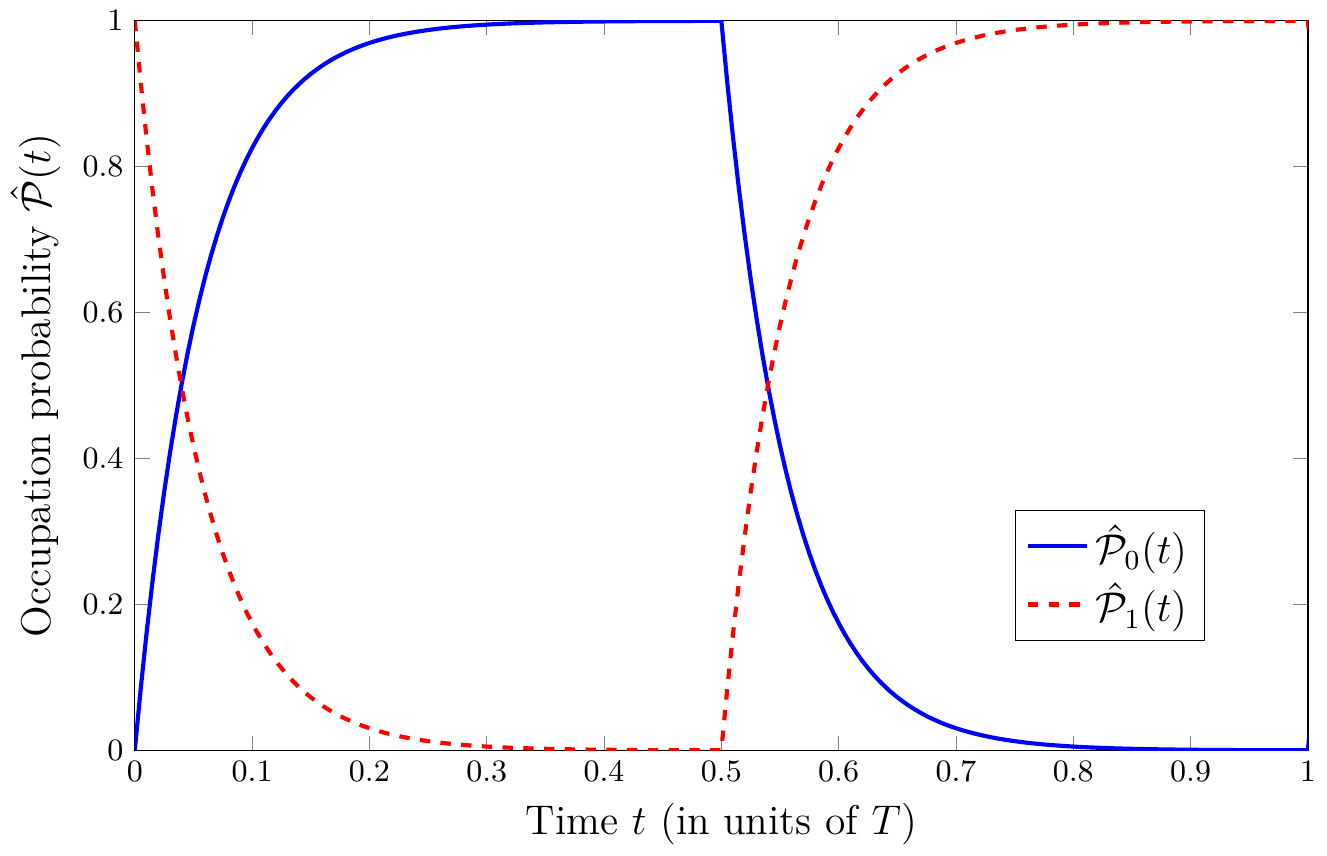}
\caption{Occupation probability of the two relevant levels of the antidot ($\hat{\mathcal{P}}_{0}$ and $\hat{\mathcal{P}}_{1}$) for the optimal regime of the SQS ($\delta = -0.2$). Other parameters are: $\nu = 1/3$,  $E_c/\omega_c = 0.01$, $\beta E_c = 20$ and $\gamma_0 T = 100$. The relevant scale entering the master equation is then $\gamma \simeq 17.4512 ~T^{-1}$.}
\label{fig3}
\end{figure}

This set of equations is simple enough to be propagated analytically. The occupation probability $\hcP_1 (t)$ is thus given by 
\begin{align}
\hcP_1 (t) &= \hcP_1 (0) e^{-\gamma t}  & \text{for~} 0 \leq t \leq \frac{T}{2} \nonumber \\
\hcP_1 (t) &= 1 + e^{-\gamma t} \left( \hcP_1 (0) - e^{\gamma T/2} \right)  & \text{for~} \frac{T}{2} \leq t \leq T.
\label{eq:solSQS}
\end{align}
while $\hcP_0 (t) = 1 - \hcP_1(t)$ at all times. The stationary state requires that $\hcP_1(0) = \hcP_1(T)$, so that $\hcP_1(0) = \frac{\exp (\gamma T/4)}{2 \cosh (\gamma T/4)}$. Numerical results for the evolution of these two probabilities over time are presented in Fig.~\ref{fig3} for a given set of parameters at filling factor $\nu=1/3$.

\subsection{Dot occupation and current}

\begin{figure}[htb]
\centering
\includegraphics[scale=0.62]{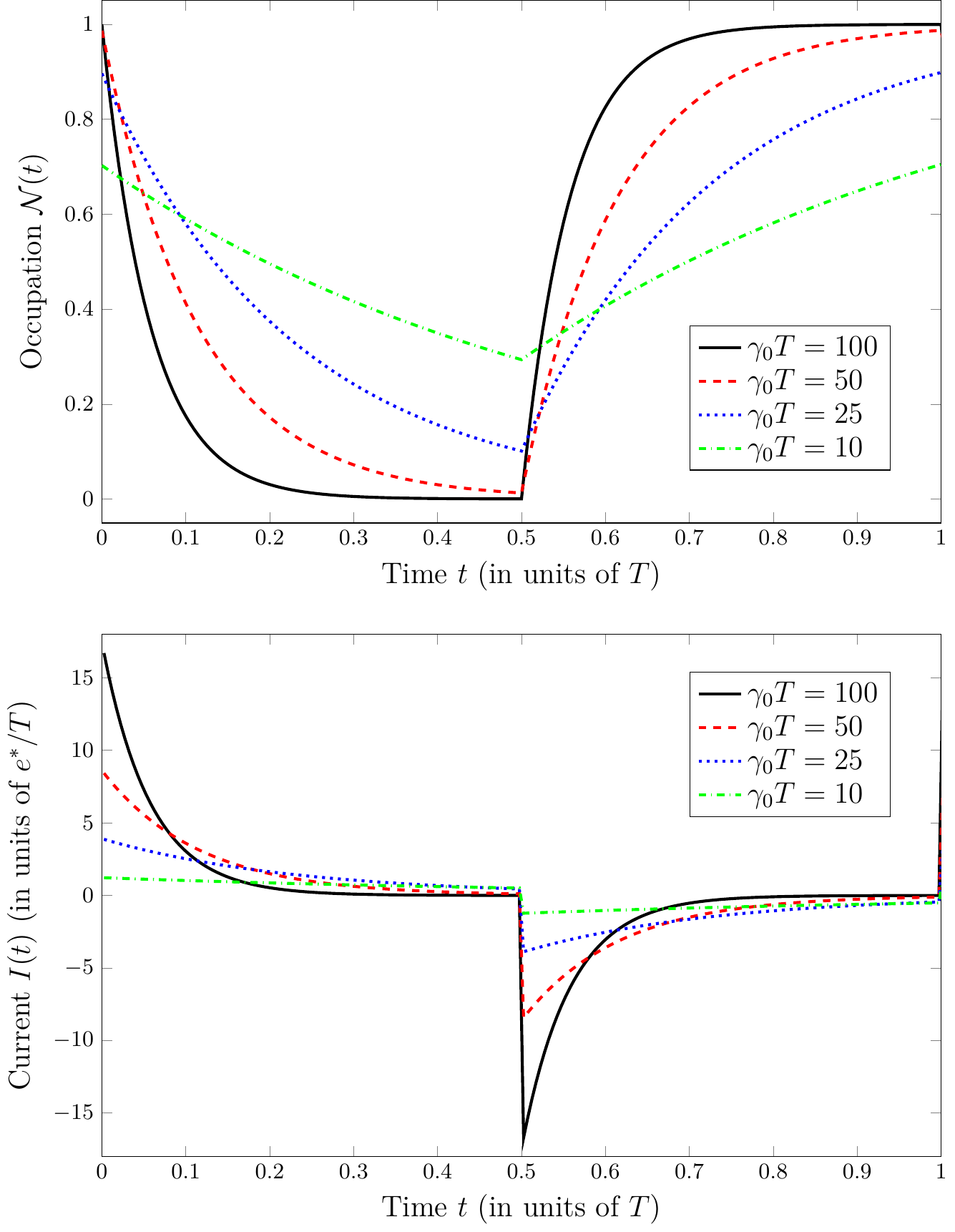}
\caption{Occupation of the antidot (top) and current flowing between the source and the edge in units of $\frac{e^*}{T}$  (bottom), in the optimal regime of SQS ($\delta=-0.2$) for different values of $\gamma_0 T$. Other parameters are: $\nu = 1/3$,  $E_c/\omega_c = 0.01$ and $\beta E_c = 20$.}
\label{fig4}
\end{figure}

The occupation of the dot is obtained directly from the occupation probabilities and Eq.~\eqref{eq:dot_occ}. Here this trivially reduces to the much simpler form
\begin{equation}
\cN (t) = \hcP_1 (t) ,
\end{equation}
and similarly for the current
\begin{equation}
I (t) = - e^* \frac{d \hcP_1 (t)}{dt} .
\end{equation}
The evolution over time of these two quantities, evaluated numerically for different values of $\gamma_0 T$, is presented in Fig.~\ref{fig4}.

Note that, because of the simple correspondence between occupation of the antidot and probability of single occupation, the escape time $\tau$ defined as the typical time associated with the emission/absorption process also corresponds to the typical time-scale governing the evolution of the two relevant occupation probabilities. In other words, one trivially has $\tau = \gamma^{-1}$.

The solution, Eq.~\eqref{eq:solSQS}, also allows to estimate the average charge transferred to the edge during every half-period (in absolute value)
\begin{align}
\overline{Q} &= e^* \left| \cN (T/2) - \cN (0) \right| \nonumber \\
&= e^* \tanh \left( \frac{\gamma T}{4} \right) .
\end{align}
confirming that the ideal operating regime is the one where $\gamma T \gg 1$ ($\overline{Q}\rightarrow e^{*}$).

\subsection{Charge fluctuations}

The time-averaged charge fluctuations can readily be derived from the computation of the conditional occupation probabilities following Eq.~\eqref{eq:cqtt}. 

\begin{figure}[ht]
\centering
\includegraphics[scale=0.62]{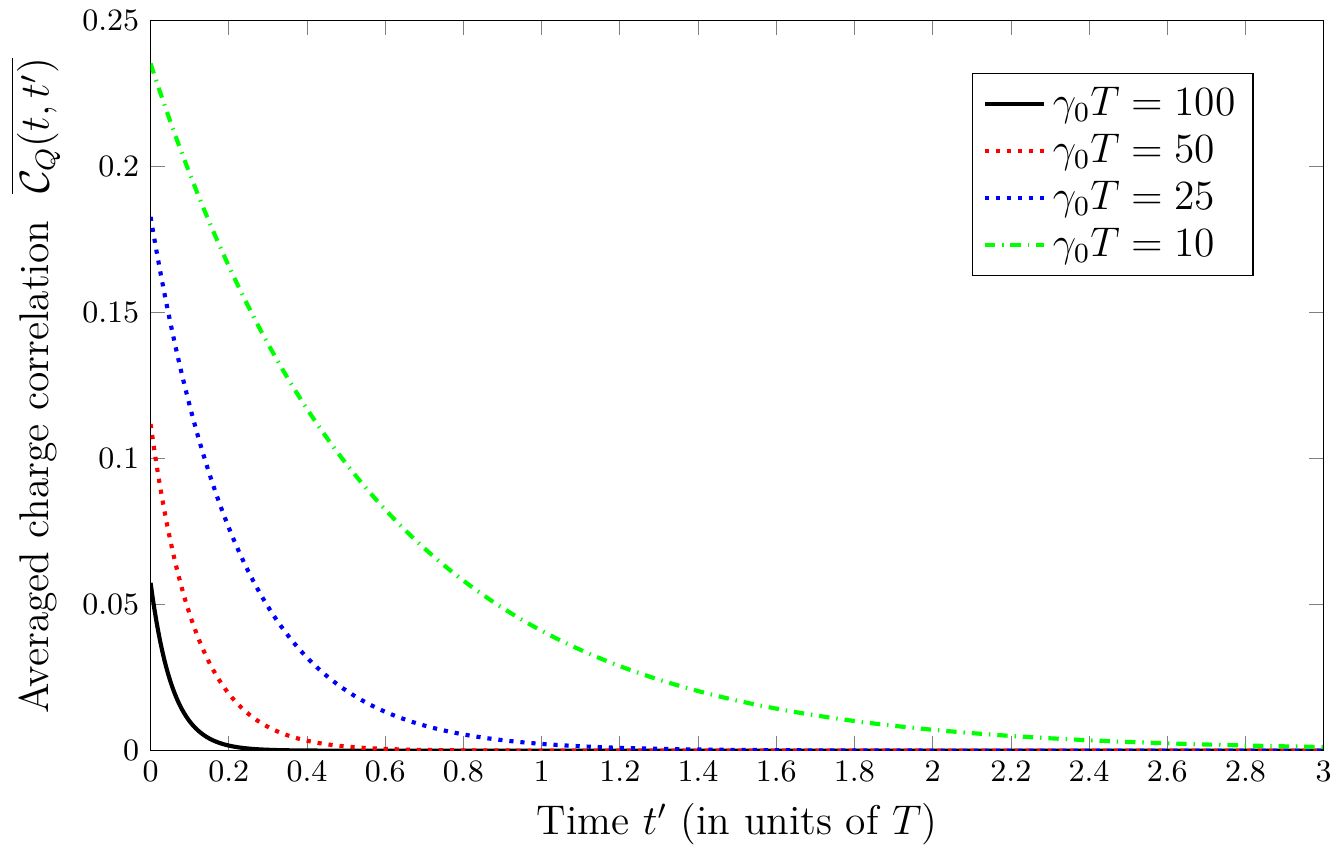}
\caption{Time-averaged charge correlation $\overline{\cC_Q (t,t')}$ in units of ${e^*}^2$, for the optimal regime of the SQS ($\delta= -0.2$) and different values of $\gamma_0 T$. Other parameters are: $\nu = 1/3$,  $E_c/\omega_c = 0.01$ and $\beta E_c = 20$.}
\label{fig5}
\end{figure}

The results are presented in Fig.~\ref{fig5}. They show that the time-averaged charge correlation $\overline{\cC_Q (t,t')}$ vanishes exponentially with the same characteristic time scale as the dot occupation, i.e. $\gamma^{-1}$. The value taken at $t'=0$ can be readily estimated from the expression for $\hcP_1 (t)$ and reads
\begin{equation}
\overline{\cC_Q (t,0)} = {e^*}^2 \frac{\tanh (\gamma T/4)}{\gamma T} .
\end{equation}
One can verify that the charge fluctuations are thus given by the following exponentially decaying form
\begin{equation}
\overline{\cC_Q (t,t')} = {e^*}^2 \frac{\tanh (\gamma T/4)}{\gamma T} e^{- |\gamma t'|} ,
\end{equation}
which is very reminiscent of what was obtained for the SES in the integer quantum Hall regime. \cite{Mahe10, Albert10, Parmentier12}

\subsection{Noise at the drive frequency}

\begin{figure}[htb]
\centering
\includegraphics[scale=0.62]{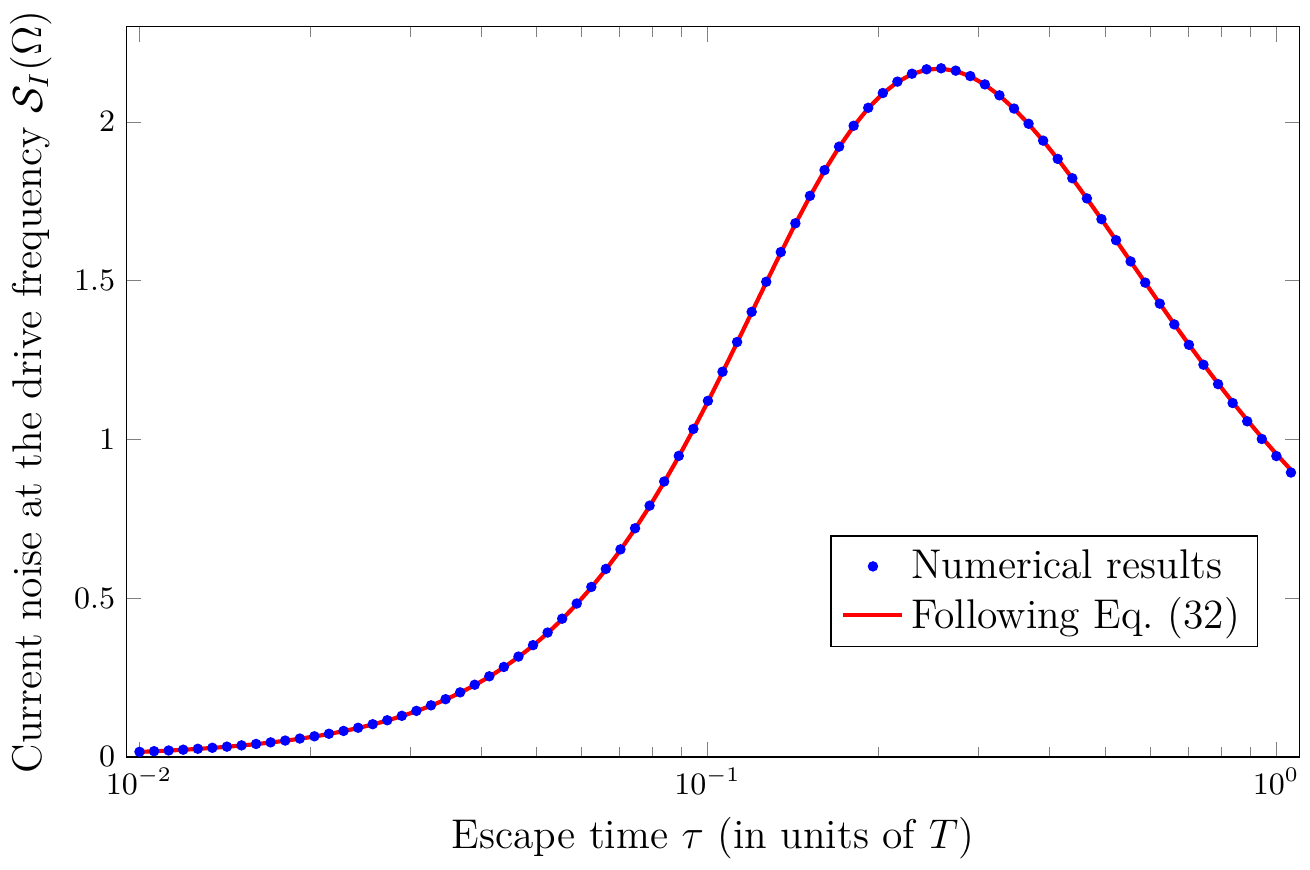}
\caption{Frequency-dependent current noise $\cS_I$ in units of $\frac{{e^*}^2}{T}$, computed at the frequency $\Omega$ of the external drive as a function of the escape time $\tau$ from the antidot, for the optimal regime of the SQS ($\delta = -0.2$). The result is compared to the analytical expectation transposed from the known results of the SES in the IQH case, assuming a fractional charge $e^{*}= \nu e$. Other parameters are: $\nu = 1/3$,  $E_c/\omega_c = 0.01$ and $\beta E_c = 20$.}
\label{fig6}
\end{figure}

Experimentally, the most accessible fluctuation-related quantity is the current noise probed at the frequency of the drive. \cite{Mahe10, Parmentier12} We computed this frequency-dependent current noise $\cS_I (\omega)$ and evaluated it at the drive frequency $\Omega = \frac{2 \pi}{T}$ for different values of $\gamma_0 T$, or equivalently different values of the escape time $\tau = \gamma^{-1}$. The results are provided in Fig.~\ref{fig6}.

Again, it makes sense here to compare the obtained behavior to the one expected for the SES in the IQH regime, as the two are governed by a similar set of equations in the present regime of operation. \cite{Albert10, Parmentier12} Not surprisingly, the current noise obtained here reproduces exactly the result of the SES upon changing $e$ into $e^*$, namely
\begin{equation}
\cS_I (\Omega) = \frac{4 {e^*}^2}{T} \tanh \left( \frac{T}{4 \tau} \right) \frac{\left( \Omega \tau \right)^2 }{1+ \left( \Omega \tau \right)^2} .
\label{eq:noiseSQS}
\end{equation}
This excellent agreement means that there are no spurious processes in the system, such as missed emissions compensated by double emissions, as well as QP/QH pairs. The source
of the observed noise for $\tau/T <1$ is the incertitude in the time of emission of the excitations during the half-period, what is usually referred to as \emph{phase} or \emph{jitter} noise. \cite{Mahe10} 
The considered emission process is therefore noiseless at long time-scales (see App. \ref{AppA}).

\section{The antidot as a SES} \label{SES}

We now turn to the case of emission/absorption of a single electron, by increasing the amplitude of the drive compared to the previous case of the SQS. Our results still rely on the numerical solution of the master equation over several drive periods, considering a vector column $\hcP$ of 31 elements ($n=-15...15$) and using a matrix version of the fourth-order Runge-Kutta scheme. 

Interestingly, this regime is much more complicated than the SQS. While one could have hoped to deal with only two occupation probabilities just as before (say $\hcP_2$ and $\hcP_{-1}$), it turns out not to be sufficient to properly describe the SES, which here cannot be reduced to a simple analytic treatment.

\begin{figure}[htb]
\centering
\includegraphics[scale=0.62]{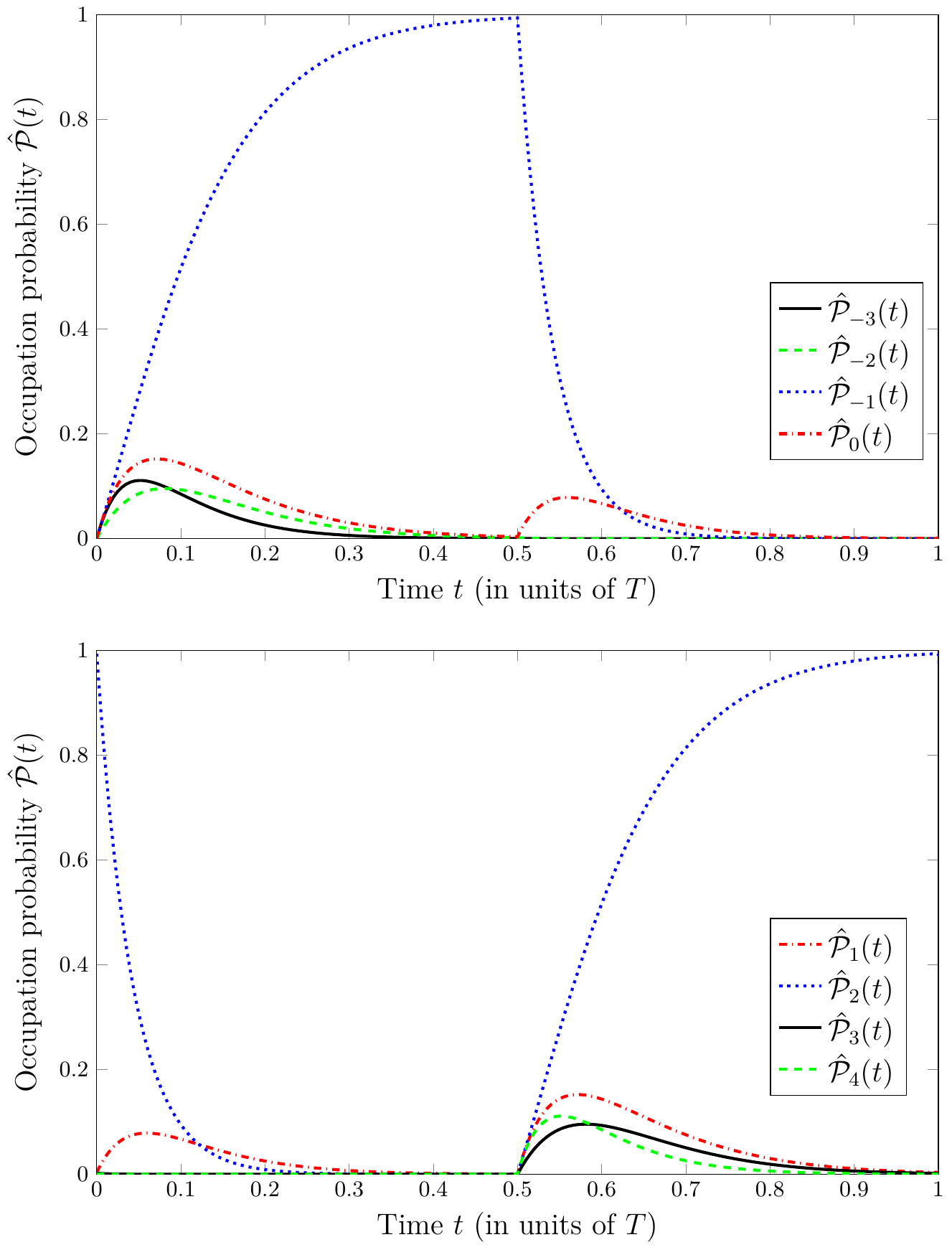}
\caption{Occupation probability of the relevant levels for the optimal regime of the SES ($\delta = -1.2$), with $\gamma_0 T = 100$. Other parameters are: $\nu = 1/3$,  $E_c/\omega_c = 0.01$ and $\beta E_c = 20$.}
\label{fig7}
\end{figure}

In practice, even in the best case scenario (very large value of $\gamma_0 T$) one needs to keep track of 8 different occupation probabilities in order to account for the behavior of the source over one drive period, and 6 if one focuses on only a half-period (the other 2 being recovered by symmetry). Numerical results for the time evolution of these 8 probabilities are presented in Fig.~\ref{fig8}, for $\gamma_0 T = 100$.
Focusing on a single half-period, corresponding e.g. to the emission of an electron, their dynamics is captured by the following set of equations
\begin{align}
\label{eq:pm3el}
\frac{d \hcP}{dt} &= \mathcal M \hcP \quad \quad \quad \quad \mbox{         with     }\\
\mathcal M  & = \left ( \begin{array}{cccccc}
             \hG_{-3,-3}  & 0 & 0 & 0 & 0 & \hG_{-3,2} \\
              \hG_{-2,-3}  & \hG_{-2,-2}  & 0 & 0 & \hG_{-2,1}  & \hG_{-2,2}\\
              \hG_{-1,-3}  & \hG_{-1,-2}  & 0 & \hG_{-1,0}  & \hG_{-1,1}  & \hG_{-1,2}\\
              \hG_{0,-3}  & \hG_{0,-2}  & 0 & \hG_{0,0}  & \hG_{0,1}  & \hG_{0,2}\\
               \hG_{1,-3}  & 0  & 0 & 0 & \hG_{1,1}  & \hG_{1,2}\\
               0 & 0 & 0 & 0 & 0 & \hG_{2,2}
             \end{array}
       \right)
\label{eq:p2el}
\end{align}
where we omitted the time dependence for notational convenience. Here, the matrix elements $\hG_{n p}$ were defined in Eq.~\eqref{eq:gammamatrix}, and we kept only the non-vanishing contributions. The other half-period, corresponding to the emission of an electron, is readily obtained from the same set of equations upon exchanging $\hcP_{-3} \leftrightarrow \hcP_{3}$, $\hcP_{-2} \leftrightarrow \hcP_{4}$, $\hcP_{-1} \leftrightarrow \hcP_{2}$ and $\hcP_{0} \leftrightarrow \hcP_{1}$ (corresponding to the exchange between similar line styles in the top and bottom panels of Fig.~\ref{fig7}).

A few comments are in order from the observation of this reduced set of equations:
\begin{itemize}
\item the apparent scattering that results in having the occupation of the antidot oscillating between 2 and $-1$ does not involve the direct exchange of 3 quasiparticles, but rather processes involving between 1 and 5 QPs,

\item during the emission process, $\hcP_2$(t) follows an exponential decay, which introduces the characteristic time-scale $\tau_2 = \hG_{2,2}^{-1}$,

\item the above set of equations can be solved by successively substituting the solution of the known probabilities into the next, according to the following order: $\hcP_{2} \to \hcP_{-3} \to \hcP_{1} \to \hcP_{-2} \to \hcP_{0} \to \hcP_{-1}$. This means in particular that all relevant occupation probabilities can be viewed as a sum of exponential terms of the form $\exp \left( \hG_{n,n} t \right)$ [note that all $\hG_{n,n}$ are negative according to Eq.~(\ref{eq:gammamatrix})].
\end{itemize}

This picture gets even more involved as one reduces the value of $\gamma_0 T$. For $\gamma_0 T = {\cal O} (1)$, about 10 occupation probabilities are required in order to properly describe the electron source, signaling the importance of processes which involve transferring up to 9 QPs.

\subsection{Dot occupation and current}

\begin{figure}[htb]
\centering
\includegraphics[scale=0.62]{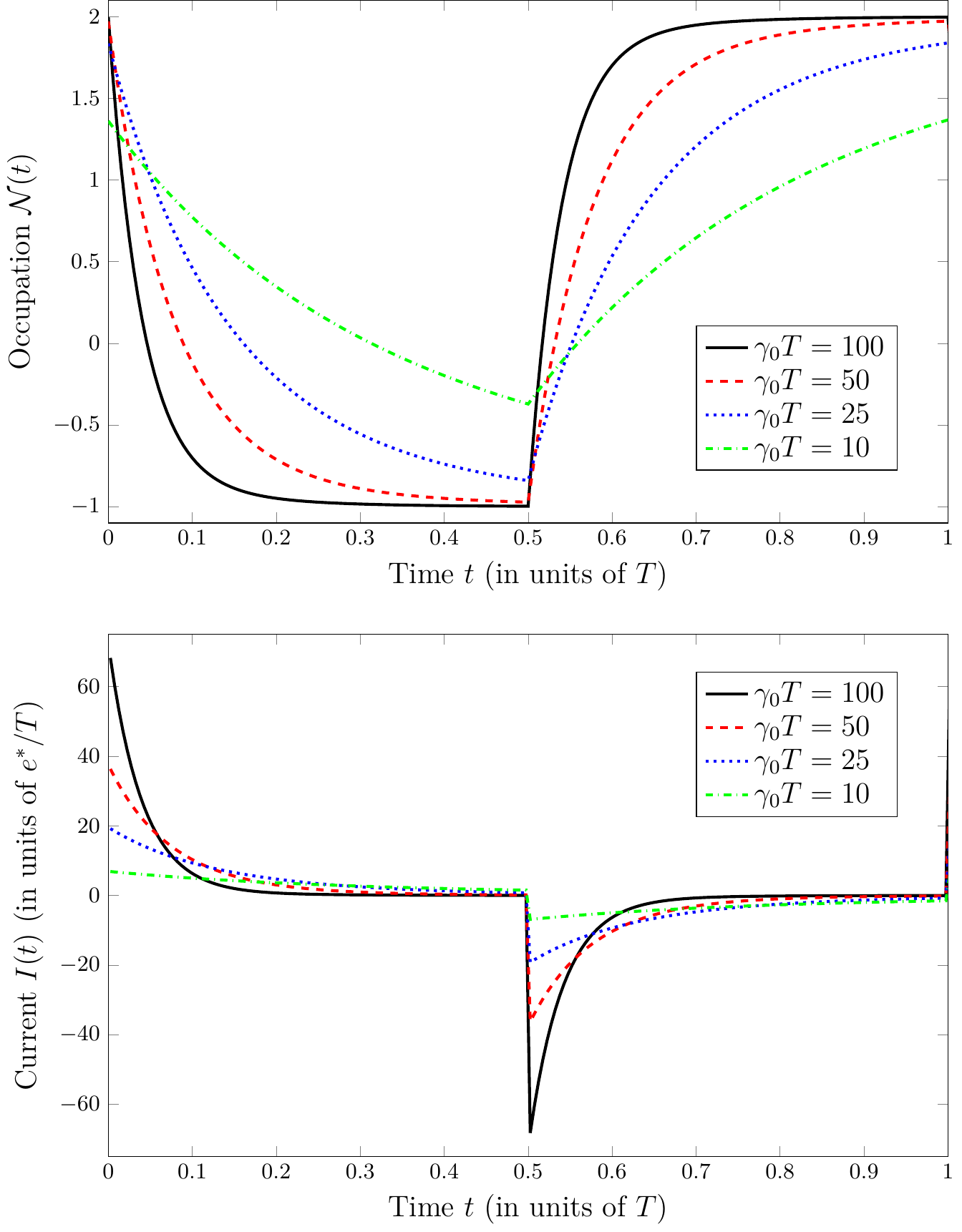}
\caption{Occupation of the antidot (top) and current flowing between the source and the edge in units of $\frac{e^*}{T}$ (bottom), in the optimal regime of SES ($\delta=-1.2$) for different values of $\gamma_0 T$. Other parameters are: $\nu = 1/3$,  $E_c/\omega_c = 0.01$ and $\beta E_c = 20$.}
\label{fig8}
\end{figure}

The occupation of the dot is obtained directly from the occupation probabilities and Eq.~\eqref{eq:dot_occ}, and similarly for the current. The evolution over time of these two quantities, evaluated for different values of $\gamma_0 T$, is presented in Fig.~\ref{fig8}. Like the individual occupation probabilities, the antidot occupation can be written as a weighted sum of exponential terms of the form $\exp (\hG_{n,n} t)$, as a direct consequence of Eq.~\eqref{eq:dot_occ}.

Note that, in contrast with the SQS, there is no proper way of unambiguously defining the escape time from the antidot. Following Eqs.~\eqref{eq:pm3el}-\eqref{eq:p2el}, one might be tempted to introduce the characteristic time-scale $\tau_{2}$ associated with the exponential decay of $\hcP_2$, the probability for single electron occupation of the antidot. However, these equations are not sufficient in the regime of low $\gamma_0 T$, and $\hcP_2$ no longer follows quite the same simple exponential decay. Similarly, one might want to define the escape time from the antidot $\tau_\text{ad}$ directly from the antidot occupation, e.g. from its initial rapid decay. However, since the occupation is a weighted sum of different exponential terms, one can only extract estimates from the early rapid decay.

A more convenient way is to follow the experimental procedure and extract the escape time from the first harmonics of the current. Indeed, one can write the first harmonics $I_\Omega = \frac{1}{T} \int_0^T dt I(t) e^{i\Omega t}$ as $I_\Omega = |I_\Omega| e^{i \phi}$ and an exponential decay of the current leads back to $\tan \phi = \Omega \tau$. Transposing this to the present quantities, one can define the escape time
\begin{align}
\tau_\Omega = - \frac{T}{2 \pi} \frac{\int_0^T dt ~\cN(t) \cos \left( \Omega t \right)}{\int_0^T dt~ \cN (t) \sin \left( \Omega t \right)}.
\end{align}

\begin{figure}[tb]
\centering
\includegraphics[scale=0.62]{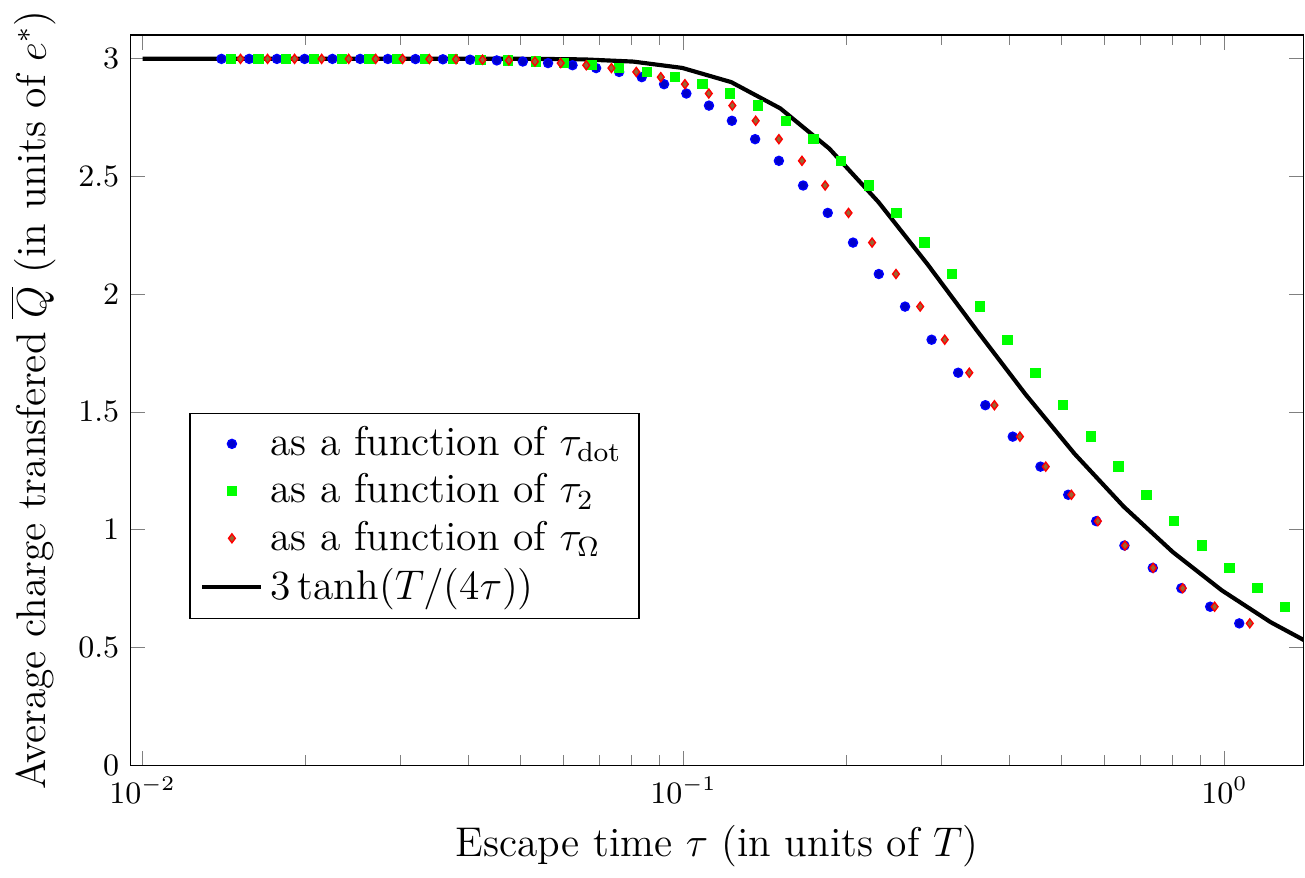}
\caption{Average charge $\overline{Q}$ transfered during one half-period, in units of $e^*$, as a function of the escape time (corresponding to different values of $\tau_\text{dot}$, $\tau_2$ and $\tau_\Omega$) and for the optimal regime of the SES ($\delta = -1.2$). Other parameters are: $\nu = 1/3$,  $E_c/\omega_c = 0.01$ and $\beta E_c = 20$.}
\label{fig9}
\end{figure}

 The average charge $\overline{Q}$ transfered to the edge during every half-period was computed numerically from the obtained solution to the master equation and is presented in Fig.~\ref{fig9}. It qualitatively behaves as in the case of the SQS, i.e. it shows a plateau at an integer value (here 3 corresponding to the fact that one electron is equal to a bunch of $3$ QPs in the Laughlin regime at $\nu=1/3$) for the lowest escape times then decreases rapidly, confirming that the ideal operating regime is the one where $\tau \ll T$. Note however that it cannot be written as simply as before, and the form $e \tanh(T/(4 \tau))$ (with any of the three definitions of $\tau$) though providing the good qualitative behavior, does not fit exactly the numerical data.

\subsection{Charge fluctuations}

\begin{figure}[b]
\centering
\includegraphics[scale=0.62]{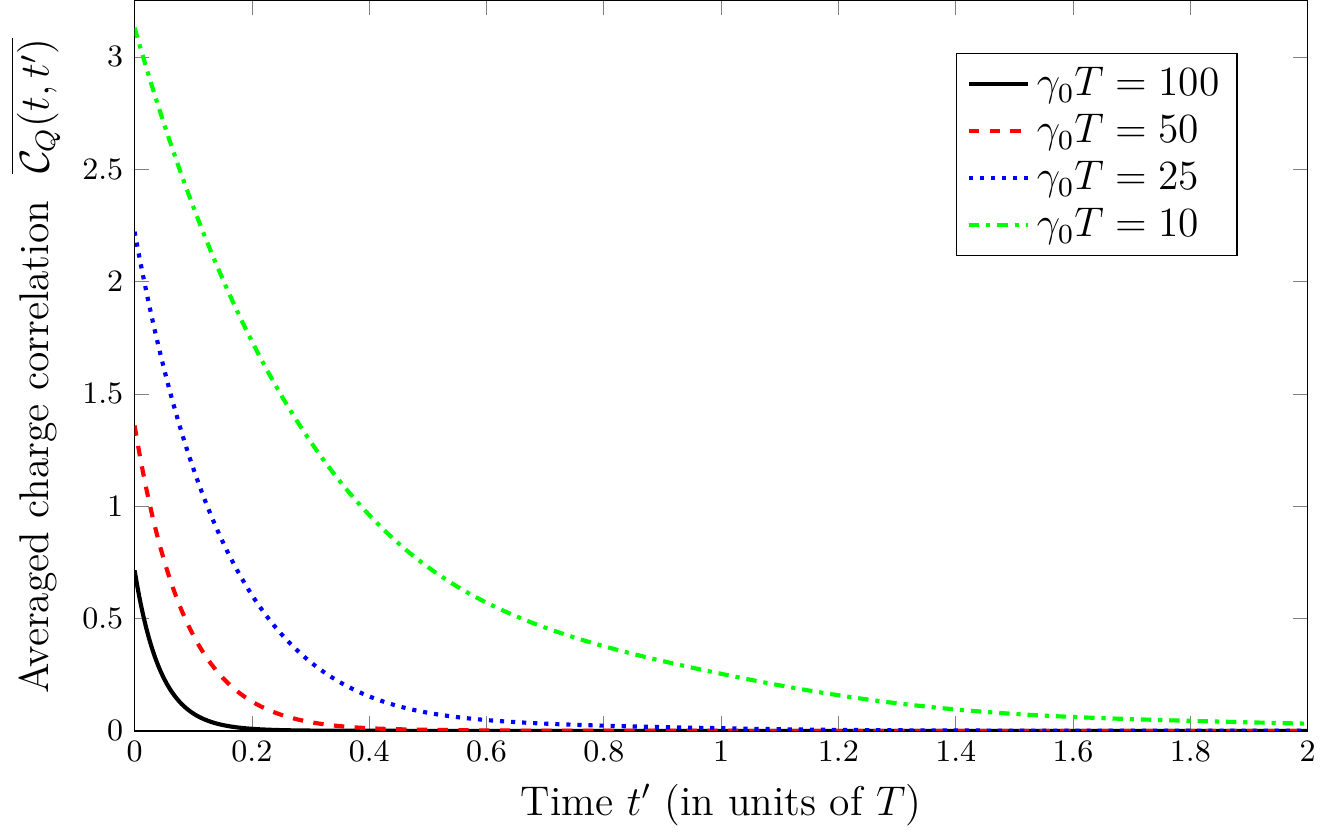}
\caption{Time-averaged charge correlation $\overline{\cC_Q (t,t')}$ in units of ${e^*}^2$, for the optimal regime of the SES ($\delta= -1.2$) and different values of $\gamma_0 T$. Other parameters are: $\nu = 1/3$,  $E_c/\omega_c = 0.01$ and $\beta E_c = 20$.}
\label{fig10}
\end{figure}

The time-averaged charge fluctuations are derived from the computation of the conditional occupation probabilities. The results are presented in Fig.~\ref{fig10}. 

As already observed in the SQS case, the time-averaged charge correlation $\overline{\cC_Q (t,t')}$ vanishes rapidly with a characteristic time-scale which seems to be similar in value to $\tau_\text{dot}$, $\tau_2$ and $\tau_{\Omega}$, without being quite exactly equal to any of those. This can be understood from Eq.~\eqref{eq:cqtt}, as $\overline{\cC_Q (t,t')}$ appears as a sum of various exponential contributions, themselves obtained by combining two different terms of the form $\exp(\hG_{n,n} t)$. 

\subsection{Noise at the drive frequency}

\begin{figure}[t]
\centering
\includegraphics[scale=0.62]{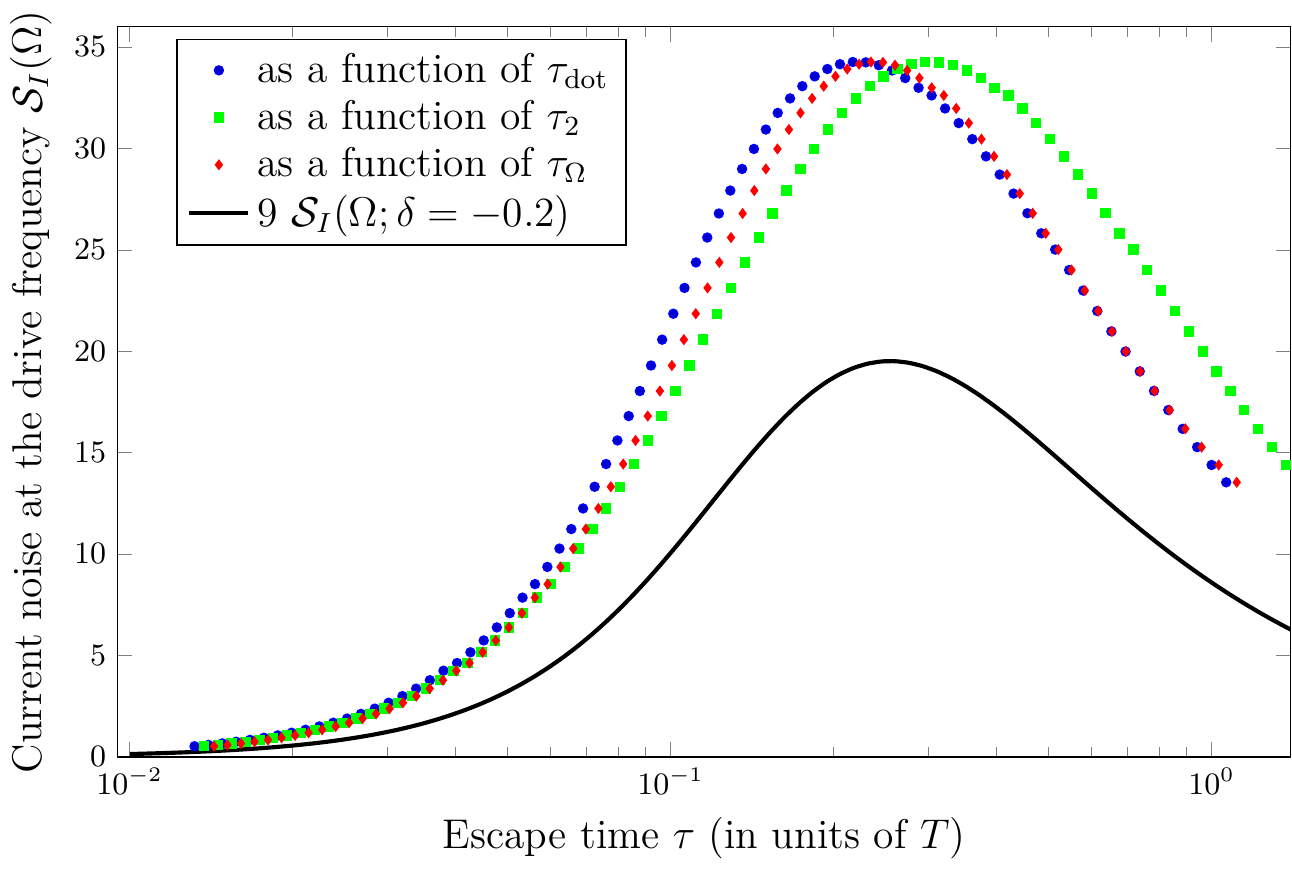}
\caption{Frequency-dependent current noise $\cS_I$ in units of $\frac{{e^*}^2}{T}$ computed at the frequency $\Omega$ of the external drive as a function of the escape time $\tau$ from the antidot, for the optimal regime of the SES ($\delta = -1.2$). Other parameters are: $\nu = 1/3$,  $E_c/\omega_c = 0.01$ and $\beta E_c = 20$. The results are compared to the analytical expectation transposed from the SQS regime, Eq.~\eqref{eq:noiseSQS}, multiplied by 9 to account for the fact that the emitted charge now corresponds to one electron (3 QPs).}
\label{fig11}
\end{figure}

We consider now the frequency-dependent current noise $\cS_I (\omega)$ associated with the SES and evaluate it at the drive frequency $\Omega = \frac{2 \pi}{T}$ for different values of $\gamma_0 T$, or equivalently different values of the escape time $\tau$ (evaluated in three separate ways: $\tau_\text{dot}$, $\tau_2$ and $\tau_\Omega$). The results are provided in Fig.~\ref{fig11}.

It makes sense here to compare the obtained behavior to the one observed in the case of the SQS, correcting for the increased charge of the carriers. Since the noise is sensible to the charge squared, we plot for comparison the result obtained in Eq.~\eqref{eq:noiseSQS} for QP injection in the previous section, multiplied by 9 (to account for the fact that the source now emits/absorbs 3 QPs per half-period). While the curves have qualitatively the same overall shape, the current noise of the SES remains much larger than the one for the SQS source, even when accounting for the increased transfered charge. This constitutes a clear indication of the importance of tunneling processes involving multiple QPs in this regime. 

Indeed, as argued when deriving the set of equations for the occupation probabilities, Eqs~\eqref{eq:pm3el}-\eqref{eq:p2el}, a proper account of the behavior of the SES requires to consider tunneling processes involving up to 9 QPs whereas we only emit an average of 3 QPs or less per half-period. Although these processes are responsible for the significant increase of the current noise at finite frequency, thus contributing to the \emph{phase} (finite frequency) noise, they do not affect the quantization of the emitted charge per half-period (see Appendix \ref{AppA}) so that the numerically evaluated frequency-dependent noise still vanishes at zero frequency.

\section{Estimating the physical parameters of the SQS} \label{sec:estimates}

For our simulations, we have chosen specific values for the various parameters which satisfy the constraints of our model, but may otherwise look random. Here we show through a simple estimate of the scales at play in an actual realization of our setup that these quantities are actually compatible with what is typically observed in electron quantum optics experiments carried out in the integer Hall case.\cite{Feve07, Mahe10, Bocquillon13b} In what follows, we reintroduce the proper factors of $\hbar$ and $k_B$ when needed.

\subsection{Energy scales}
\label{scales}

Focusing on the state at filling factor $\nu=1/3$, assuming a small antidot of circumference $L\approx 1$ $\mu$m \cite{Franklin96} and a propagation velocity along the edge $v\approx 10^{5}$ m/s, one has the charging energy 
\be
E_{c}/k_{B}\approx 0.8 \text{K}
\ee
(expressed in units of temperature) which is smaller, but of the same order of magnitude as the energy gap observed in the SES in the IQH regime ($\Delta\approx 1.4$ to $4.2$ K).  \cite{Feve07, Bocquillon13a, Bocquillon13b} 
The plasmonic energy is larger
\be
\epsilon/k_{B}\approx 4.8 \text{K},
\ee
but plays only a marginal role in the dynamics of the device.

The temperature of the system, as assumed in the simulations, is 
\be
\Theta= 0.05 E_{c}/k_{B}\approx 40  \text{mK}.
\ee
It is comparable to the one obtained in experiments involving the antidot geometry \cite{Franklin96, Goldman01} and not out of reach from  nowadays experimental techniques which allow to carry out electron quantum optics measurements ($\Theta\approx 60$ to $100$ mK). \cite{Bocquillon13a, Bocquillon13b}

The optimal operation regime of the SQS discussed in the text is reached for a drive amplitude $\delta = -0.2$, which corresponds to a magnetic field fluctuation of
\be
\Delta B= 0.2 ~\frac{4 \pi \Phi_{0}}{L^{2}}\approx 10^{-2} \text{T}.
\ee 
This represents a small and experimentally achievable fluctuation with respect to the magnetic field needed to realize the FQH states \cite{Goldman01, Goldman05} (typically of the order of $10\text{T}$, depending on the two dimensional electron density of the considered sample).

The high-frequency cutoff $\omega_{c}= v/\alpha$ represents the highest energy scale in the model. We assume that the short length cutoff $\alpha$ is typically set by the magnetic length of the system, so that  $\alpha\approx 10$ nm. This in turn leads to $\omega_{c} \approx 10$ THz, or equivalently
\be
\hbar \omega_{c}/k_B \approx 76 \text{K}
\ee
from which one readily sees that $E_c/\omega_c \simeq 0.01$ as used in our simulations.

\subsection{Time scales}

In order to operate the antidot in the optimal regime of emission, the typical time-scales of the device need to satisfy the relation 
\be
\tau_{0} < \gamma^{-1} < \frac{T}{2}
\label{time_constraints}
\ee
where $\tau_{0}=L/v$  is the time required to make a loop around the antidot. 
The first inequality ensures the validity of the continuous limit assumed in the master equation approach, as $\tau_{0}$ corresponds to the discretization time associated with the semiclassical processes. \cite{Albert10} The second one guarantees that the emission of a QP (QH) is achieved during one half-period. According to the previous estimation one has 
\be
\tau_{0}=\frac{L}{v}\approx 10 \text{ps}.
\ee

Assuming a driving frequency $f=500$ MHz, close to the GHz regime investigated in Refs. \onlinecite{Feve07, Mahe10}, the drive period is given by 
\be
T= \frac{1}{f}\approx 2 \text{ns}
\ee 
so that the half-period is of the order of the nanosecond.

This enforces the emission time from the SQS to vary in the range $10$ to $1000\text{ps}$, compatible with what is currently observed in similar experiments ($\gamma^{-1}\approx 60$ to $900\text{ps}$ in the optimal regime for the SES in the IQH case).\cite{Feve07, Bocquillon13a, Bocquillon13b} Note that any value of the emission time $\gamma^{-1}$ in this range also satisfies the condition
\be
\hbar \gamma < k_{B} \theta < E_{c}
\ee
necessary to fulfill the sequential tunneling approximation.\cite{Beenakker91, Furusaki98}  

The maximum value of the emitted current is directly related to the emission time, as one has
\begin{equation}
I_\text{max} = e^* \gamma \frac{\exp(\gamma T/4)}{2 \cosh (\gamma T/4)}
\end{equation}
which corresponds to a value of the current between $10\text{pA}$ and $1\text{nA}$, a reasonable range in comparison with what is experimentally measured for the SES in the integer regime\cite{Feve07} and for the continuous current measurement in the weak backscattering regime at fractional filling factor.\cite{Chung03} 

Finally, using our energy-scale estimates, we can propagate these bounds on the emission time and obtain
\begin{equation}
10^{-4} < |t_{L}|^{2} < 10^{-2},
\end{equation}
for the range of transmission amplitude between the antidot and the nearby edge channel.

\section{Conclusions} \label{Conclusions}

In this paper we have analyzed a strongly asymmetric antidot geometry realized through depletion of the Hall fluid, and periodically driven in time by a modulated magnetic flux. Through a master equation approach, we discussed the possibility to use this kind of setup both as a SQS and a SES. In the first case, only two charge states of the antidot are involved and the dynamics of the system allows for a tractable analytic treatment. It shows the quantization of a fractional charge $e^{*}=\nu e$ emitted during each half-period and fluctuations analogous to the ones observed for the electron emission in the integer quantum Hall case. The electron emission regime proves more complicated as various charge states of the antidot are involved, requiring a full numerical treatment. 

Here, we observe that we can reach a regime where the emitted charge over a half-period is precisely $e$ (with the precise opposite charge emitted over the second half of the period).
However, the noise measured at the drive frequency is strongly enhanced with respect to what is observed in the integer regime due to the random emissions of additional excitations having zero mean charge (quasiparticle-quasihole pairs, etc.) which provide an essential contribution to the dynamics of the system at finite frequency. 
In spite of the presence of these additional tunneling events, it is possible to extract information about the escape time of the electron by looking at the first current harmonic as is usually carried out in experiments.      

It is worth underlying that in previous works, discussing similar geometries realized with Luttinger liquids \cite{Braggio03} (typically in CNT \cite{Cavaliere04} or two dimensional topological insulators \cite{Dolcetto13}) the role of the magnetic flux is played by an external gate voltage. The possibility to exploit a fluctuation of the same gate voltage used to realize the antidot also to induce the QP injection could be fruitful from the experimental point of view, an external electrostatic gate being easier to tune than a magnetic field. However, in this case, modifications of the antidot geometry are expected and their effects on the functionality of the presented device have to be carefully taken into account. 

The present study could prove quite relevant for the implementation of both theoretical and experimental investigations of interferometric phenomena in  HBT\cite{HBT} and HOM\cite{HOM} setups in the context of  electron quantum optics. Indeed, the present SQS or SES source could be embedded in a quantum Hall bar which is divided in two by a central quantum point contact, where either the HBT partitioning of the source could be analyzed in order to quantify the production of spurious excitations, or alternatively with two antidot sources where two QP collisions could be achieved in order to probe the overlap of QP wave packets. In the near future, we are determined to model such experiments assuming that either ideal QP wavepackets or ideal electron wavepackets have been deposited on the fractional Hall edge,\cite{rech_fqhe} but a proper description of the source, such as presented in the context of this work, will ultimately be necessary to bring the description sufficiently close to experimental reality. 

Furthermore, the present study was achieved assuming the weak coupling hypothesis where tunneling rates are smaller than the electronic temperature. The numbers we provided in Sec. \ref{scales} seem to point out that this hypothesis is justified in the context of present experimental working conditions ($k_B\Theta\simeq 50mK$). However, upon either increasing the tunneling rates (by bringing the antidot closer to the edge, which would increase the QP tunneling amplitude) or by working at much lower temperatures, a coherent description of tunneling will be eventually required. This constitutes a truly challenging task, which may have to rely on Keldysh non equilibrium Green's function formalism in order to describe the time dependence of the current, while taking into account the (finite) QP occupation of the antidot in its evolution.  

\acknowledgements

Early discussions with D. C. Glattli are gratefully acknowledged.
The authors also acknowledge the support of Grants No. ANR-2010-BLANC-0412 (``1 shot'') and No. ANR-2014-BLANC (``one shot reloaded''). 
This work was granted access to the HPC resources of Aix-Marseille Universit\'e financed by the project Equip@Meso (Grant No. ANR-10-EQPX-29-01) and has been carried out in the framework of the Labex ARCHIMEDE (Grant No. ANR-11-LABX-0033) and of the AMIDEX project (Grant No. ANR-11-IDEX-0001-02), all funded by the ``investissements d'avenir'' French Government program managed by the French National Research Agency (ANR).

\appendix 
\section{Charge fluctuations during a period} \label{AppA}
This Appendix discusses the relation between the current noise at zero frequency and the fluctuations of the charge emitted by a SQS or a SES during one period.  Let us start by considering the general definition of the noise 
\be
S(t, t') = \langle I(t) I(t')\rangle - \langle I(t)\rangle \langle I(t')\rangle
\label{general_noise}
\ee
where $I$ is the current flowing along the considered channel and the averages are taken with respect to an arbitrary initial state. 

The operator associated with the charge emitted during one period $T$ is given by \cite{Chtchelkatchev02}
\be
Q(T)= \int^{+T/2}_{-T/2} d \tau I(\tau).
\ee

Together with Eq.~(\ref{general_noise}), the above definition leads to 
\begin{align}
\langle Q(T) Q(T) \rangle &= \int^{+T/2}_{-T/2} dt dt' \langle I(t) I(t')\rangle \nonumber\\ 
&= \int^{+T/2}_{-T/2} dt  dt' S (t, t')+\langle Q(T) \rangle \langle Q(T) \rangle. 
\label{eq:QTQT}
\end{align}

Taking into account the periodicity of the noise signal and using the standard change of variables \cite{Ferraro13} 
\beq
t&=& \bar{t}+\frac{\tau}{2}\\
t'&=&\bar{t}-\frac{\tau}{2}, 
\eeq
the first term in Eq.~\eqref{eq:QTQT} becomes 
\begin{align}
\int^{+T/2}_{-T/2} dt  dt' S (t, t') 
&= \int^{+T/2}_{-T/2} d\bar{t} \int^{+\infty}_{-\infty} d\tau S \left( \bar{t}+\frac{\tau}{2}, \bar{t}-\frac{\tau}{2}\right)\nonumber\\
&= \int^{+T/2}_{-T/2}  d\bar{t} \int^{+\infty}_{-\infty}   d\tau  \sum_{m=-\infty}^{+\infty}   \frac{1}{T}  \nonumber \\ 
&\quad \times  \int^{+\infty}_{-\infty}   \frac{ d \omega}{2 \pi} e^{-i m \frac{2 \pi}{T} \bar{t}} e^{-i \omega \tau} \tilde{S}^{(m)}(\omega) \nonumber \\ 
&= \tilde{S}^{(m=0)}(\omega=0),
\end{align}
where in the second line we introduced the Fourier transform associated with the variable $\tau$ and the series relative to $\bar{t}$, while in the last line the only remaining term is given by the $m=0$ harmonic of the noise evaluated at zero frequency.

Summarizing the above results, one finds 
\be
\langle Q(T) Q(T) \rangle=  \tilde{S}^{(m=0)}(\omega=0)+\langle Q(T) \rangle \langle Q(T) \rangle.
\ee
Concerning the cases considered in this paper, analytical and numerical evidences suggest that 
\be
\tilde{S}^{(m=0)}(\omega=0)\approx 0 
\ee
and consequently that
\be
\langle Q(T) Q(T) \rangle-\langle Q(T) \rangle \langle Q(T) \rangle \approx 0.
\ee
This is a clear signature of the absence of charge fluctuations during one period for both the SQS and the SES realized in the fractional regime, in full analogy with what is observed for the SES in the integer quantum Hall case. \cite{Mahe10} However, charge fluctuations due to the uncertainty in the moment of QP emission from the source still remain. This so-called \emph{phase} or \emph{jitter} noise provides an intrinsic finite frequency contribution\cite{Bocquillon13b} which can be detected by measuring the noise at the frequency of the drive.

\end{document}